
%
%
\magnification=1200
\baselineskip=24pt
\overfullrule=0pt
\rightline{IMSc--92/56}
\rightline{December 15, 1992}
\vskip .6in
\centerline{\bf Complete Solution of $SU(2)$ Chern-Simons Theory}

\vskip .8in

\centerline{R.K. Kaul \footnote*{email: kaul@imsc.ernet.in}}
\centerline{The Institute of Mathematical Sciences}
\centerline{Taramani, Madras 600 113, India}

\vskip .5 in

\noindent {\bf Abstract}

     Explicit and complete topological solution of $SU(2)$  Chern-Simons theory
on $S^3$ ~is presented.

\vskip 1 in

\vfill
\eject

     Chern-Simons theories have been of immense interest in recent
times$^{1-7}$. These find application in many areas of physics$^{2,3}$.
Pure Chern-Simons theories are also closely related to knots
and links$^{4-7}$. Knot theory is relevant, for example, in
physics of polymers as well as to the study of some properties
of biological molecules$^8$.

     Chern-Simons action is given by
$$kS~=~ {k \over 4\pi} \int\limits_{S^3} tr (AdA~+~{2 \over 3}~A^3)
\eqno(1)$$

\noindent where A is the matrix valued connection one-form of gauge group
$SU(2)$~. The topological operators of this theory are the Wilson
link operators. For a link L made of oriented knots ~$(C_1,C_2,....C_s)$~
carrying ~$SU(2)$~ spins ~$j_1,j_2....j_s$~ on them respectively, the
Wilson link operator is
$$W_{j_{1}j_{2}...j_s}[L]~=~\prod\limits_{l=1}^s~tr_{j_l}~P~ exp~ \oint_{C_{l}}
A
\eqno(2)$$

\noindent Here the one-form A carries matrix representations
corresponding to respective spin $j_l$~ on the knot ~$C_l$~. We are
interested in the expectation value of these operators:
$$V_{j_1j_2...j_s}~[L]~=~Z^{-1}~ \int ~[dA]~ W_{j_1j_2...j_s}
[L]~ e^{ikS} ,
\quad\quad Z~~=~~ \int [dA] ~e^{ikS} \eqno(3)$$

     The partition function is given by $^4$ : $Z=[2/(k+2)]^{1/2}
{}~sin(\pi/(k+2))$~. Placing doublet representations on all components,
Witten has also shown that these link invariants satisfy the
same skein relations as those by Jones polynomials$^9$. These
skein relations can be recursively solved to obtain Jones polynomials
for an arbitrary link. Generalization of skein relations to the cases
other than spin half have also been obtained$^{6,7}$. But unfortunately
these cannot be solved recursively to obtain the expectation values (3)
for an arbitrary link. In refs. 7 we developed a direct method to
obtain such invariants for links which are related to 4-strand braids.
But links which cannot be so constructed stayed elusive. In the following
we present a method which is applicable to arbitrary links in $S^3$. To do so
we shall make use of theory of coloured-oriented braids and the duality
properties of correlators of $SU(2)_k$ Wess-Zumino conformal field
theory on an $S^2$.

     An $n$-braid is a collection of $n$ non-intersecting strands
connecting two parallel rigid rods. When the strands are oriented and
coloured, we have
coloured-oriented braids. We shall colour the strands by putting ~$SU(2)$~
spins on them. A general braid can thus be specified by giving $n$
assignments, $\hat j_i~=~(j_i,\epsilon_i)$~ representing the
spin $j_i$~ and orientation $\epsilon_i(= \pm1$ for the strand going into or
away from the rod
) on the $n$ points where the strands meet the
upper rod and also $n$ spin-orientation assignments $\hat l_i =
(l_i,\eta_i)$ on the $n$ points on the lower rod. This braid will be
represented symbolically as ${\cal B}_n \left(\matrix{\hat j_1 &\hat j_2...
&\hat j_n \cr
\hat l_1&\hat l_2...&\hat l_n\cr}\right)$. For a given spin-orientation
assignment $\hat j_i~=~(j_i, \epsilon_i)$
, we define the conjugate assignment as
$\hat j^*_i~=~(j_i, -\epsilon_i)$. Then assignments
$(\hat l_i)$ are just a permutation  of $(\hat j^*_i)$.

	 An arbitrary
$n$-braid can be generated by
applying braiding
generators $B_i, i=1,2,...$ $n-1$ on the ``identity'' braids $I_n \left(\matrix
{\hat j_1 &\hat j_2...&\hat j_n \cr
\hat j_1^*&\hat j_2^*...&\hat j_n^*\cr}\right)$. These are represented in
fig. 1.
There are more than one identity braid. Further, composition between two
braids is defined only if there is colour-orientation matching along the
rods that are merged. These braids form a groupoid instead of a group.
The generators, however, still satisfy the same generating relations
as  for the ordinary braids. Also platting construction of ordinary
braids$^{10}$
can also be extended to coloured-oriented braids.
Consider a $2m$-braid with special spin-orientation assignments
as ${\cal B}_{2m} \left(\matrix
{\hat j_1 &\hat j_1^* &\hat j_2 &\hat j_2^*...&\hat j_m &\hat j_m^*\cr
\hat l_1&\hat l_1^* &\hat l_2 &\hat l_2^*...&\hat l_m &\hat l_m^*\cr}\right)$.
Platting then constitutes of pairwise joining the the successive
strands $(2i-1,2i), i=1,2,... m$ from above and below the two rods.
There is a theorem due to Birman for ordinary  braids which
relates links with plats.$^{10}$.  This theorem can obviously also
be stated for our coloured-oriented braids:

 	$\underline {\rm Proposition~ 1}$:~~ A coloured-oriented knot or link
can be represented (though not
 uniquely) by a platting of an oriented-
coloured braid ${\cal B}_{2m} \left(\matrix
{\hat j_1 &\hat j_1^* ...&\hat j_m &\hat j_m^*\cr
\hat l_1&\hat l_1^* ...&\hat l_m &\hat l_m^*\cr}\right)$.

     In addition to this property, we need
one more ingredient for  our discussion. This is the duality properties
of correlators of ~$SU(2)_k$~ Wess-Zumino conformal field theory on an
{}~$S^2$~. For example, the four-point correlators for primary fields
with spins $j_1,j_2,j_3,j_4$ (such that these four spins form a singlet) can
be represented in three different equivalent ways. Two such ways are
shown in figs. 2(a, b). The trivalent points in these diagrams
all satisfy fusion rules which for sufficiently large $k$ are
same as the triangle conditions for $SU(2)$~ spins. These two linearly
independent sets of conformal blocks, $\phi_j(j_1j_2j_3j_4)$~ and
$\phi^\prime_l(j_1j_2j_3j_4)$~ are related by duality$^{11,7}$:
$$\phi_j(j_1j_2j_3j_4)~=~\sum\limits_l
a_{jl} \left[\matrix{j_1&j_2 \cr
j_3&j_4\cr}\right]~~ \phi^\prime_l~(j_1j_2j_3j_4) \eqno(4)$$
where duality matrices are explicitly given in terms of $SU(2)$~ quantum
Racah coefficients$^{12}$.

$$
a_{jl}~\left[\matrix{j_1&j_2 \cr
j_3&j_4}\right]
{}~=~(-)^{j+l+min(j_1+j_3,j_2+j_4)} \sqrt{[2j+1][2l+1]}
\left(\matrix{j_1&j_2&j \cr
j_3&j_4&l \cr}\right)\eqno(5a)$$
$$\quad\quad\quad\eqalign{\left(\matrix{j_1&j_2&j_{12} \cr
j_3&j_4&j_{23} \cr}\right)~=~& \bigtriangleup~(j_1j_2j_{12})~
\bigtriangleup~(j_3j_4j_{12})~\bigtriangleup~(j_1j_4j_{23})~
\bigtriangleup ~(j_3j_2j_{23}) \cr
&\times \sum_{m\geq 0}~(-)^m[m+1]!\{ [m-j_1-j_2-j_{12}]!
[m-j_3-j_4-j_{12}]! \cr
&\times [m-j_1-j_4-j_{23}]!
[m-j_3-j_2-j_{23}]!
[j_1+j_2+j_3+j_4-m]!\cr
&\times [j_1+j_3+j_{12}+j_{23}-m]![j_2+j_4+j_{12}+j_{23}-m]!\}
^{-1}
 \quad\quad}\eqno(5b)$$

$$\bigtriangleup~(abc)~=~ \sqrt{{[-a+b+c]![a-b+c]![a+b-c]! \over
[a+b+c+1]!}} \eqno(5c)$$

\noindent Here the square brackets represent the q-numbers,
$[x] = (q^{x/2}-q^{-x/2})/(q^{1/2}-q^{-1/2})$ with $q = exp
(2\pi i/(k+2))$.

     This duality relation can be extended to arbitrary $2m$-point
correlators. In particular we shall be interested in
the correlators $\phi_{(p;r)} (j_1...j_{2m})$ and $\phi^\prime_{(q;s)}
(j_1...j_{2m})$ represented in figs. 3(a, b) for primary fields
carrying spins $j_1j_2...j_{2m}$. We have used a compact notation
for spins on the internal lines in these diagrams as $p = (p_0p_1...p_{m-1}),
{}~r=(r_1r_2...r_{m-3})$ and similarly $q = (q_0 q_1...q_{m-1}),
{}~s=(s_1s_2...s_{m-3})$. These two
figures represent two equivalent ways of combining $2m$ spins
$j_1j_2...j_{2m}$~ into singlets and are related by duality:
$$\phi' _{(q;s)}~(j_1j_2...j_{2m})~=~\sum\limits_{(p;r)}~
a_{(p;r)(q;s)} \left[ \matrix {j_1&j_2 \cr
j_3&j_4 \cr
\vdots &\vdots \cr
j_{2m-1}&j_{2m} \cr} \right]
\phi_{(p;r)} (j_1j_2...j_{2m})
\eqno(6a)$$

The duality matrices here are given in terms of those given in
eqns.5 for four-point correlators :
$$a_{(p;r)(q;s)}~\left[ \matrix{j_1&j_2 \cr
\vdots&\vdots \cr
j_{2m-1}&j_{2m} \cr} \right]
{}~=~ \sum\limits_{t_1,t_2..t_{m-2}}
{}~~\prod\limits^{m-2}_{i=1}~
(a_{t_ip_i}~ \left[ \matrix{r_{i-1}&j_{2i+1} \cr
j_{2i+2}&r_i \cr} \right]
{}~~a_{t_is_{i-1}}~ \left[ \matrix{t_{i-1}&q_i \cr
s_i&j_{2m} \cr} \right]) $$

$$\quad\quad\quad\quad\quad\quad\quad\quad
\quad\quad\quad\times ~\prod^{m-2}_{l=0}~~a_{r_l q_{l+1}}~~\left[ \matrix
{t_l&j_{2l+2} \cr
j_{2l+3}&t_{l+1} \cr} \right].
\eqno(6b)$$

\noindent where $r_0 \equiv p_0, r_{m-2} \equiv p_{m-1}, ~t_0 \equiv
{}~j_1, t_{m-1} \equiv j_{2m}, s_0\equiv q_0, s_{m-2} \equiv q_{m-1}$ and
${\vec j}_1 + {\vec j}_1+... + {\vec j}_{2m-1} = {\vec j}_{2m}$
and spins meeting at the trivalent points in figs.3 satisfy the triangle
relations. The proof of this statement can be developed by repeated
applications of the duality transformation involving four points
given by  eq.(4).

     Now we are in a position to
develope the solution of the $SU(2)$ Chern-Simons theory.
Let us consider an $S^3$ from which two 3-balls have been
removed. This is a manifold with two boundaries, each an $S^2$. Let
$2m (m=1,2,3,....)$ Wilson lines carrying spin-orientations
$\hat j_1\hat j_2.... \hat j_{2m}$ without
any entanglements connect them as shown in fig.4(a). Thus we have placed
the identity braids $I_{2m}$ inside this manifold.
The Chern-Simons functional integral over this manifold,
following Witten$^4$ , can be thought of as a state in the tensor product of
the vector spaces, ${\cal H}^{(1)} \otimes {\cal H}^{(2)}$, associated
with the two boundaries.
These vector spaces are related
to the space of conformal blocks for $2m$-point correlators with spin
assignments
$(j_1j_2....j_{2m})$ of the $SU(2)_k$ Wess-Zumino
theory on these boundaries. Corresponding to the conformal blocks shown in
figs. 3(a and
b), we have two possible sets of basis vectors for each of these
vector spaces, $\mid \phi_{(p;r)}(\hat j_1\hat j_2
... \hat j_{2m})>$ and $\mid \phi^\prime_{(q;s)}(\hat j_1\hat j_2
...  \hat j_{2m})>$ respectively. The
corresponding bases for the dual vector spaces associated with oppositely
oriented boundaries will be represented by $<\phi^{(p;r)}(\hat j_1
\hat j_2....  \hat j_{2m}) \mid$ and $<\phi^{\prime (q;s)}(\hat j_1
\hat j_2....  \hat j_{2m}) \mid$ respectively. Gluing two
manifolds along two such oppositely oriented boundaries represents a
natural
product of the vectors. The bases vectors under this product
are normalized so that
$$<\phi^{(p;r)}(\hat j^*_1
\hat j^*_2....  \hat j^*_{2m}) \mid
\phi_{(p^\prime ;r^\prime)}(\hat j_1
\hat j_2....  \hat j_{2m}) > = \delta_{(p,p^\prime)}
{}~\delta_{(r,r^\prime)}$$
$$<\phi^{\prime (q;s)}(\hat j^*_1
\hat j^*_2....  \hat j^*_{2m}) \mid
\phi^\prime_{(q^\prime ;s^\prime)}(\hat j_1
\hat j_2....  \hat j_{2m}) > = \delta_{(q,q^\prime)}
{}~\delta_{(s,s^\prime)}  \eqno(7)$$
\noindent Further, the primed and unprimed bases for each of the vector
spaces are related by the duality matrices of eqns.6:
$$\mid ~ \phi^\prime_{(q;s)}~(\hat j_1
\hat j_2....  \hat j_{2m}) > ~=~
\sum\limits_{(p,r)}~a_{(p;r)(q;s)}~~\left[\matrix {\hat j_1 &\hat j_2 \cr
\hat j_3 &\hat j_4 \cr
\vdots &\vdots \cr
\hat j_{2m-1} &\hat j_{2m} \cr} \right]
{}~~\mid \phi_{(p;r)} ~(\hat j_1
\hat j_2....  \hat j_{2m}) >
\eqno(8)$$

     Now the Chern-Simons functional integral over the
three-manifold of fig. 4(a) can thus be expanded in terms of one of
these bases  for each boundary as
$$\nu_I~=~\sum\limits_{(p;r)}~~\mid~\phi^{(1)}_{(p;r)}~
(\hat j^*_1 \hat j^*_2 .... \hat j^*_{2m}
)>~~ \mid ~\phi^{(2)}_{(p;r)} ~(\hat j_1 \hat j_2 ....
 \hat j_{2m})>
\eqno(9)$$
\noindent Here we have put superscripts (1) and (2) explicitly on the
basis vectors to indicate that these belong to the vector spaces
${\cal H}^{(1)}$ and ${\cal H}^{(2)}$ associated with the two
boundaries respectively.

     The conformal blocks shown in fig. 3 and therefore the corresponding
bases $\mid \phi_{(p;r)}>$ and $\mid \phi^\prime_{(q;s)} >$
are eigen functions of the
odd-indexed braid generators $B_{2l+1},\quad\quad l = 0,1, .... m-1$ and
even-indexed generators, $B_{2l}, l = 1,2,.... m-1$, respectively:
$$B_{2l+1} \mid \phi_{(p;r)} (\hat j_1 \hat j_2 ... \hat j_{2l+1}
\hat j_{2l+2}... \hat j_{2m})> ~=~\lambda_{q_{l}} (\hat j_{2l+1},
\hat j_{2l+2}) \mid \phi_{(p;r)}~(\hat j_1 \hat j_2 ... \hat j_{2l+2}
\hat j_{2l}... \hat j_{2m})>$$
$$B_{2l} \mid \phi^\prime_{(q;s)} (\hat j_1 \hat j_2 ... \hat j_{2l}
\hat j_{2l+1}... \hat j_{2m})> ~=~\lambda_{q_{l}}(\hat j_{2l},
\hat j_{2l+1}) \mid \phi^\prime_{(q;s)}~(\hat j_1 \hat j_2 ... \hat j_{2l+1}
\hat j_{2l}... \hat j_{2m})> \eqno(10)$$
\noindent The eigenvalues of these  half-twist matrices depend on the
relative orientation of the strands involved $^{7,11}$
$$\lambda_t (\hat j, \hat j^\prime) ~=~ \lambda^{(+)}_t
(j,j^\prime) \equiv
{}~(-)^{j+j^\prime-t} q^{(c_j+c_{j\prime})/2 + c_{m in (j,j^\prime)}
- ~{c_t/2}}~~\quad\quad if~~ \epsilon \epsilon^\prime = + 1$$
{}~~~~~~~~~~~~~~~~~~~~~~~~~~~~~~~~~$$=~(\lambda^{(-)}_t (j,j^\prime))^{-1}
\equiv
{}~(-)^{ \mid j-j^\prime\mid-t} q^{ \mid c_j-c_{j^\prime} \mid
/2-c_t/2~~}\quad\quad\quad \quad if ~~
 \epsilon \epsilon^\prime = - 1 \eqno(11)$$
\noindent where $c_j = j(j+1)$.

     These braiding generators can be applied to the identity braids
of fig.(4a) to obtain a general braid inside the manifold.
Thus the  braid represented by the shaded box in fig.(4b) can be
written as a word ${\cal B}$ in terms of these generators. Using eqn.9 we
can represent the functional integral over this three-manifold
as:
$$\nu_{\cal B}~~=~~~~ \sum\limits_{(p;r)} ~\mid \phi^{(1)}_{(p;r)}> {\cal B}
\mid \phi^{(2)}_{(p;r)}> \eqno(12)$$

    To plat this braid
consider the Chern-Simons
functional integral over the ball shown in fig. 4c. This functional
integral can again be represented by a vector in the vector space
associated with the boundary.
It is proportional to the basis vector $\mid \phi_{(0;0)} (\hat j_1 \hat
j_1^*....
\hat j_m \hat j_m^*)>$ which is the eigen function of the odd indexed
braiding generators
with eigenvalue 1. Further gluing two copies
of this manifold onto each other along oppositely oriented boundaries
yields $m$ untangled unknots.
The invariant for these is simply the product of invariants
for individual unknots$^7, [2j_i+1],~ i=1,2...m$. Thus the functional
integral (normalised by multiplying by $Z^{-1/2}$) for the 3-ball
of fig. 4c is
$$\nu~ =( ~\prod
^m_{i=1} [2j_i+1]^{1/2}) ~\mid \phi_{(0;0)} ~(\hat j_1 \hat j_1^*\hat j_2 \hat
j_2^*....\hat j_m \hat j_m^*)>
\eqno(13)$$

     Now we are ready to plat the braid shown in fig. 4(b) by gluing
it from above and below by two copies of 3-ball of fig. 4c along
oppositively oriented boundaries with spin-orientation assignments
matching properly. This leads us to our main result which we now state:

	 $\underline {\rm Proposition~ 2}$: The expectation value of
a Wilson operator for a link L obtained by platting an oriented-coloured
$2m$-braid ${\cal B}_{2m} \left(\matrix {\hat j_1&\hat j_1^*&\hat j_2&\hat
j_2^*
..&\hat j_m&\hat j_m^* \cr
\hat l_1&\hat l^*_1&\hat l_2&\hat l^*_2
..&\hat l_m&\hat l^*_m \cr} \right)$
represented by a word in terms of the braid generators is given by
$$V[L]~=~(\prod\limits^m_{i=1}[2j_i+1])~<
\phi^{(0;0)}(\hat l^*_1 \hat l_1...
\hat l^*_m \hat l_m) \mid {\cal B}_{2m}
{}~~\left(\matrix {\hat j_1 &\hat j_1^*
..&\hat j_m&\hat j_m^* \cr
\hat l_1 &\hat l_1^*
..&\hat l_m&\hat l_m^* \cr} \right)
{}~~\mid \phi_{(0;0)}
(\hat j^*_1 \hat j_1...\hat j^*_m \hat j_m)> \eqno(14)$$
	\noindent Propositions 1 and 2  along with the fact that two bases $\mid
\phi_{(p;r)} >$
and $\mid \phi^\prime_{(q;s)} >$ are related by the duality matrices
as given by eqns. (8) and (6b) above  allow us to
calculate explicitly the functional average (3) for any arbitrary link.
This provides the complete topological solution  of the $SU(2)$~ Chern-Simons
theory
on $S^3$. The method has obvious generalization to other gauge groups
as well as other three-manifolds.

     Placing spin $1/2$ representation on all the component knots
of a link, gives us the Jones polynomial. Placing other representations
on the knots yields a whole variety of new invariants. These
invariants are more powerful than Jones invariant as these do distinguish
knots which are represented by the same Jones polynomial. The
discussion of these aspects as well as more elaborate details of the
proofs above will be presented elsewhere.

\noindent$\underline {\bf References}$

\noindent 1. A.S. Schwarz, Lett. Math. Phys. $\underline 2$ (1978) 217;
G.V. Dunne, R. Jackiw and C.A. Trugenberger, Ann. Phys. (N.Y.)
$\underline {194}$(1989)197; M.Bos and V.P. Nair, Phys. Lett. $\underline
{B223}$ (1989) 61;
S. Elitzur et al.Nucl. Phys.
$\underline {B326}$ (1989) 108.

\noindent 2. Y. Chen, et al. Int. J. Mod. Phys. $\underline {B3}$
(1989) 1001.

\noindent 3. A.M. Polyakov, Mod. Phys. Lett. $\underline {A3}$ (1988) 325.

\noindent 4. E. Witten, Commun. Math. Phys. $\underline {121}$ (1989) 351.

\noindent 5. M. Atiyah, The Geometry and Physics of Knots, Cambridge
Univ. Press (1990).

\noindent 6. K. Yamagishi, M-L Ge and Y.S. W, Letts. Math. Phys.
$\underline {19}$ (1990) 15;
Y. Wu and K. Yamagishi, Int. J. Mod. Phys. $\underline {A5}$
(1990) 1165;
 J.H. Horne, Nucl. Phys.$\underline{ B334}$ (1990) 669; S. Mukhi, TIFR preprint
TIFR/TH/89-39.

\noindent 7. R. K. Kaul and T. R. Govindarajan, Nucl. Phys.
$\underline{B380}$(1992) 293; Chern-Simons theory as a theory of knots and
links II: multicoloured
links, Nucl. Phys. B (in press); Rama Devi, T. R. Govindarajan and R. K. Kaul,
IMSc. preprint IMSc/92/55.

\noindent 8. K. Koniaris and M. Muthukumar, Phys. Rev. Lett. $\underline
{66}$ (1991) 2211.

\noindent 9. V. F. R. Jones, Bull. AMS $\underline {12}$ (1985) 103;
Ann. of Math. $\underline {128}$ (1987) 335.

\noindent 10. J.S. Birman, Braids, Links and Mapping Class Groups, Annals of
Mathematics Studies, Princeton Univ. Press (1975).

\noindent 11. G. Moore and N. Seiberg, Lectures on RCFT in ``superstring 89'',
ed. M. Green et al World Scientific, Singapore (1990);
L. Alvarez Gamme and G. Sierra, CERN preprint TH 5540/89.

\noindent 12. A.N. Kirillov and N. Yu. Reshetikhin, in New Developments
in the Theory of Knots, Ed. T. Kohno, World Scientific, Singapore
(1989).

\vfill
\eject

$\underline {\rm Figure~Captions}$

\noindent Fig. 1.   Identity braids and braid generators

\noindent Fig. 2.   Duality transformation of 4-point correlators

\noindent Fig. 3.   Two equivalent sets of conformal blocks for
2m-point correlators

\noindent Fig. 4.   Functional integrals over manifolds with boundaries
\end